# Long-range nonlocal flow of vortices in narrow superconducting channels

I. V. Grigorieva, A. K. Geim, S.V. Dubonos and K.S. Novoselov
Department of Physics and Astronomy, University of Manchester, Oxford Road, Manchester M13 9PL, UK

D. Y. Vodolazov and F. M. Peeters
Departement Natuurkunde, Universiteit Antwerpen, Universiteitsplein 1, B-2610 Antwerpen, Belgium

P. H. Kes and M. Hesselberth
Kamerlingh Onnes Laboratorium, Leiden University, P.O. Box 9504, 2300 RA Leiden, The Netherlands

We report a new nonlocal effect in vortex matter, where an electric current confined to a small region of a long and sufficiently narrow superconducting wire causes vortex flow at distances hundreds of inter-vortex separations away. The observed remote traffic of vortices is attributed to a very efficient transfer of a local strain through the one-dimensional vortex lattice, even in the presence of disorder. We also observe mesoscopic fluctuations in the nonlocal vortex flow, which arise due to "traffic jams" when vortex arrangements do not match a local geometry of a superconducting channel.

Phenomena associated with vortex motion in type-II superconductors have been subject to intense interest for many decades, as they are important both for applications and in terms of interesting, complex physics involved. Vortices start moving when the driving (Lorentz) force acting on them exceeds pinning forces arising from always-present defects. The Lorentz force $f_L$ is determined by the local current density $j$ and, hence, the resulting vortex motion is confined essentially to the region of a superconductor where the applied current flows [1,2]. There are only a few cases known where vortex flow becomes nonlocal (i.e. not limited to the current region), most notably in Giaever's flux transformer [3] and in layered superconductors [4]. In the former case, $f_L$ is applied to vortices in one of the superconducting films comprising the transformer, while the voltage is generated in the second film, due to electromagnetic coupling between vortices in the two films [3,5]. In layered superconductors, the electric current can be distributed non-uniformly along the anisotropy axis, and a drag effect (somewhat similar to that in Giaever's transformer) is observed due to coupling between pancake vortices in different layers. Both nonlocal effects occur along vortices and are basically due to their finite rigidity. A high viscosity of a correlated vortex matter can also lead to a nonlocal response in the direction *perpendicular* to vortices [6-9]. In this case, local vortex displacements induced by $j$ create secondary forces on their neighbours pushing them along. Such nonlocal correlations have been observed in the vicinity of the melting transition in high-temperature superconductors [8,9]. This is a dynamic effect where regions of the vortex lattice – generally moving at different speeds due to different above-critical currents – suddenly become locked in a long-range collective motion. In the absence of a driving current, such viscosity-induced nonlocality is expected to die off at a few vortex separations [6,7] and has not been observed so far.

In this paper, we report a nonlocal effect of a different kind, which arises *in the absence of a driving current* due to a long-range collective response of a rigid one-dimensional (1D) vortex lattice and survives at strikingly long distances, corresponding to several hundreds vortex spacings. Nonlocal vortex flow in our experiments is observed at distances up to ≈5 μm, provided a superconducting channel contains only one or two vortex rows. To the best of our knowledge, such nonlocality has neither been observed nor even considered in theory. We propose a simple model that explains the observed signal and is supported by numerical simulations using Ginzburg-Landau (GL) equations.

Starting samples in our experiments were thin films of amorphous superconductor MoGe ($\kappa \approx 60$) with various thicknesses $d$ from 50 to 200 nm. We have chosen amorphous films because they are known for their quality and very low pinning and have been extensively studied in the past in terms of pinning and vortex flow (see, e.g., [10,11]). The sharp superconducting transitions (<0.1 K) measured on mm-sized samples of our films indicate their high quality and homogeneity. The critical current $j_C$ in intermediate fields $b = H/H_{c2} \approx 0.3 - 0.6$ was measured to be $\approx 10^2$ A/cm$^2$ (at 5K), where $H$ is the applied field and $H_{c2}$ the upper critical field. $j_C$ increased several times at lower temperatures. The MoGe films were patterned into multi-terminal submicron wires of various widths $w$ (between 70 nm and 2 μm) and lengths $L$ (between 0.5 and 12 μm), using e-beam lithography and dry etching. An example of the microfabricated structures is shown in Fig. 1. Electrical measurements were carried out using the standard low-frequency (3 to 300 Hz) lock-in technique at temperatures $T$ down to 0.3 K. The results were independent of frequency, which proves that the measured ac signals are just the same as if one were using a dc measurement technique, provided the latter could allow the same sensitivity (< 1nV). The external field $H$ was applied perpendicular to the structured films. For brevity, we focus below on the results obtained in the nonlocal geometry and omit discussions of the complementary measurements and characterization carried out in the standard (local) four-probe geometry.

The nonlocal geometry is explained in Fig. 1. Here, the electric current is passed through leads marked $I^+$ and $I^-$ and voltage is measured at terminals $V^+$ and $V^-$. In this geometry, the portion of applied current $I$ that goes sideways along the central wire (see Fig. 1) and reaches the area between the voltage probes is negligibly small. Indeed, in both the normal and superconducting states [12], the current along the central wire decays as $\propto I \cdot \exp(-\pi x/w)$, which means that the current density reduces by a factor of ten already at distances $x \approx w$ and, typically, by $10^{10}$ in the nonlocal region ($x = L$) in our experiments. This also means that all vortices in the central wire, except for one or two nearest to the current-carrying wire, experienced the current density many orders of magnitude below the critical value. Therefore, no voltage can be expected to be observable in the nonlocal geometry.

In stark contrast, our measurements revealed a pronounced nonlocal voltage $V_{NL}$, which emerged just below [13] the critical temperature $T_C$ and persisted deep into the superconducting state (Fig. 1). The signal appeared above a certain value of $H \approx 0.2 H_{c2}$, reached its maximum at $(0.5\text{-}0.7) \cdot H_{c2}$ and then gradually disappeared as $H$ approached $H_{c2}$. $V_{NL}$ was found to depend linearly on $I$ that was varied between 0.2 and 5 µA. At lower $I$, $V_{NL}$ became so small (< 100pV) that it disappeared under noise, while higher currents led to heating effects. The linear dependence allows us to present the results in terms of resistance $R_{NL} = V_{NL}/I$. With increasing $L$, $R_{NL}$ was found to decay relatively slowly (for $L \le 4$ µm) and quickly disappeared for longer wires as well as for the wide ones ($w \ge 0.5$ µm) (Fig.2). The general shape of $R_{NL}(H)$-curves was identical for all samples but fluctuations (sharp peaks) seen in Fig. 1 varied from sample to sample. They were however reproducible for consecutive cooldowns. A closer inspection of the fluctuations for different samples shows that they have the same characteristic interval of magnetic field over which $R_{NL}$ changes rapidly. This correlation field $B_C$ corresponds to the entry of one flux quantum $\Phi_0$ into the area $L \cdot w$ between the current and voltage leads, so that $B_C \approx \Phi_0/L \cdot w$. For example, in Fig. 1 $B_C$ is about 200 G, in agreement with the wire dimensions (1 µm long, ≈150 nm wide).

To understand the nonlocal signal, it is important to note that within the accessible range of $I$, its density inside the current-carrying wire was in the range of $\approx 10^3$ to $10^5$ A/cm$^2$ (i.e. $\gg j_C$) and, accordingly, caused a vortex flow through this wire. Indeed, whenever $V_{NL}$ was observed, measurements in the local geometry showed the behaviour typical for the flux flow regime. This indicates that the nonlocal resistance is related to the vortex flow in the current-carrying part of the structures, which then somehow propagates along the central wire to the region between $V^+$ and $V^-$ terminals, where no electric current is applied. The mechanism of the propagation can be understood as follows. The Lorentz force – acting on vortices located at the intersection between current-carrying and central wires – pushes/pulls them along the central wire. In the absence of edge defects along this wire, the surface barrier prevents these vortices from leaving a superconductor [14] and, hence, the local distortion of the vortex lattice can be expected to propagate along the central wire, away from the current-carrying region. If the vortex motion reaches the remote intersection between the central and voltage wires, a nonlocal voltage is generated by vortices passing through this region. For an infinitely rigid vortex lattice, such a local distortion would of course propagate any distance. However, for a soft vortex lattice and in the presence of some disorder, the lattice can be compressed and vortices become jammed at pinning sites. The softer the vortex lattice, the shorter the distance over which the distortion is dampened. Note that, as we discuss a dc phenomenon, there should exist a constant flow of vortices through the sample. We believe that this is ensured by large contact regions that inevitably have weak points for vortex entry/exit along their extended edges. The regions then act as vortex reservoirs.

The interplay between pinning and elasticity of a vortex lattice is important in many vortex phenomena, and the spatial scale, over which a vortex lattice behaves as almost rigid (responds collectively), is usually determined by Larkin-Ovchinnikov correlation length $R_C$ [15,16]. This concept had been successfully used in the past to explain the behaviour of $j_C$ in macroscopic thin films, where the only relevant elastic modulus defining $R_C$ is the shear modulus $C_{66}$ [17,18]. For our particular films, the maximum value of $R_C$ can be estimated as ≈ $20a_0$ (reached at ≈$0.3H_{c2}$) and then $R_C$ gradually reduces to ≈ $a_0$ as $H$ approaches $H_{c2}$ (here, $a_0 \approx (\Phi_0/B)^{1/2}$ is the vortex lattice period and $B$ the magnetic induction) [10,19]. This length scale is in agreement with theoretical predictions [6,7] and clearly too short to explain the observed $R_{NL}$. For example, at 4.5 K, $R_{NL}$ was detected at distances up to 5 μm and in fields up to 3.5 T. This means that the entire vortex ensemble between the current and voltage wires, which is over 200 vortices long, is set in motion by a localised current.

To explain these unexpectedly long-range correlations, we argue that the vortex lattice in mesoscopic wires is much more rigid than in macroscopic films due to its 1D character and the presence of the edge confinement that prevents transverse vortex displacements. Indeed, if there are only a few vortex rows in a narrow channel, the only possible deformation of the lattice is via uniaxial compression. This deformation is described by compressional modulus $C_{11} >> C_{66}$. In this case, the characteristic length, over which one should expect collective response, is much longer and given by another correlation length $\lambda_C = (C_{11}/\alpha_L)^{1/2}$, where $\alpha_L \approx F_p/r_p$ is a characteristic of the pinning strength, $F_p = j_C \cdot B$ the bulk pinning force, and $r_p$ the pinning range ($r_p \approx a_0/2$ for $b$ >0.2) [20,21,2]. It is interesting to note that $\lambda_C$ is given by the same formula as the Campbell length [20], which describes the penetration of ac field in bulk superconductors (although the corresponding expressions for $C_{11}$ are different because of different dimensionalities). However, the main difference is that, in the case of ref. [20], vortices collectively oscillate in a *static* lattice around their equilibrium positions, while in our experiments they actually move over large distances ≈ $\lambda_C$. The difference is clearly emphasized by the fact that the observed $R_{NL}$ exists only in 1D samples that contain no more than a few vortex rows.

To calculate $\lambda_C(H)$ we used the expression $C_{11} \approx \Phi_0 \cdot B/(2 \cdot \mu_0 \cdot \lambda^2 \cdot a_0 \cdot k)$ expected for a 1D channel [22] (not the standard expression for bulk superconductors [23]). Here, $\lambda$ is the field- and temperature- dependent penetration depth [22,23] and $k$ the wave-vector of vortex lattice deformation. Our numerical simulations show that the most relevant $k$ is given by the lattice distortion in the cross-shaped regions (see Fig. 3(b)) and, accordingly, we assume $k \approx 1/w$. The estimated $\lambda_C$ in intermediate fields at $T = 6K$ is ≈ 3 - 10 μm, in agreement with our experiment. The above model also describes well the observed field dependences of $R_{NL}$. The theory curve in Fig. 3(a) takes into account that the nonlocal signal should decay as $R_{NL} \propto \exp(-L/\lambda_C)$ where $\lambda_C = (\Phi_0 \cdot w)^{1/2}/2\lambda(\mu_0 \cdot j_C)^{1/2}$ and that, for narrow wires, it is thermodynamically unfavourable for vortices to penetrate the narrow wires until $H$ reaches a critical value $H_S \approx \Phi_0/\pi\xi w$. The latter effect is modelled by pinning at the surface barrier, which results in an additional part in $j_C \propto \exp(-H/H_S)$. The disappearance of $R_{NL}$ below 4 K is attributed to stronger pinning (higher $j_C$) at lower $T$. Note that the exponential dependence implies that changes in $j_C$ by a factor of 4, which occur below 5 K, result in a rapid suppression of $R_{NL}$.

It is clear that the above description applies only to wires that accommodate just a few vortex rows. As the number of rows increases, the vortex lattice gains an additional (lateral) degree of freedom and a local compression becomes dampened by both longitudinal and lateral deformations. One eventually expects a transition to the 2D case described by the shear modulus $C_{66}$ and a much shorter correlation length $R_C$. In addition, as more vortex rows are added, elastic correlations are expected to become less relevant, as vortex dynamics becomes dominated by plastic deformation of the vortex lattice [22]. The latter is forbidden in a 1D case but in wider channels it can become a dominant mechanism for dampening of collective flow. This qualitatively explains the disappearance of $R_{NL}$ in wider wires.

To support the discussed model further, we carried out direct numerical simulations of nonlocal vortex traffic, using time-dependent GL equations. Details of the analysis will be published elsewhere and here we use these results to elucidate the physics behind our observations. The middle curve in Fig. 3(a) plots a typical example of the obtained field dependence of $R_{NL}$ for the geometry shown in Fig. 3(b). One can see that the GL simulations reproduce the overall shape of $R_{NL}(H)$ observed experimentally. Furthermore, the numerical analysis allowed us to clarify the origin of fluctuations in $R_{NL}(H)$: they appear due to sudden changes in vortex configurations. Fig. 3(b) shows such changes for the field marked by the arrow in Fig. 3(a), where a sharp fall in $R_{NL}$ is observed. Here, approximately two additional vortices enter the central wire, which results in a transition from an easy-flow vortex configuration ($H = 0.52 H_{c2}$) to a blocked one ($0.56 H_{c2}$). In the latter case, vortices in the cross-shaped regions are distributed rather randomly and break down the continuity of the vortex rows formed in the central wire. This leads to blockage of collective vortex motion. For $H = 0.52 H_{c2}$, vortices in the cross' areas are more equally spaced, and the corresponding vortex rows make a shallower angle with rows in the central wire (Fig.3(b)). In this case, there is less impediment to vortex motion through the cross regions which leads to a larger nonlocal voltage.

The mechanism of the sudden blocking/unblocking of vortex flow at different $H$ becomes even clearer if one considers an imaginary configuration containing just a few vortices – see Fig. 3(c). Here, we find a sharp fall in $R_{NL}$ when the number of vortices changes from 9 to 11 ($N = 10$ is a thermodynamically unstable state for this geometry). For $N = 9$, the vortex row passes continuously through the whole central wire, allowing its motion as a whole when pushed/pulled along by a localised current. In contrast, for $N = 11$, there is a vortex pair in each of the crosses which prevents such vortex motion.

In conclusion, we have studied nonlocal resistance in narrow superconducting wires and observed pronounced flux flow at distances corresponding to hundreds of vortex lattice periods from the region where applied current flows. We attribute the observed behaviour to an enhanced rigidity of the vortex lattice confined in narrow channels and provide a theoretical model for this. Our results also show that by confining a vortex system to a long narrow channel one can manipulate vortices by a remote force, which we believe can be used for development of new fluxtronics devices.

Acknowledgments. We thank M. Blamire, M. Moore and V. Falko for helpful discussions. The work was supported by EPSRC (UK), Ministry of Industry and Science (Russia) and the Flemish Science Foundation.

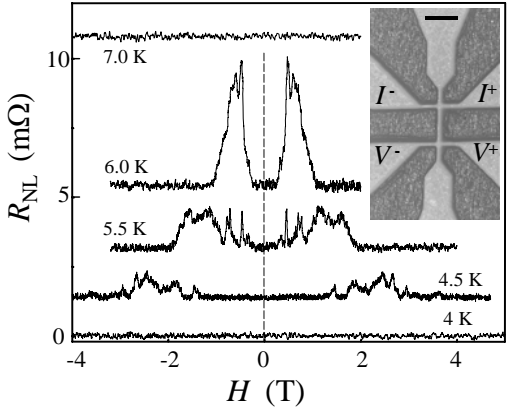
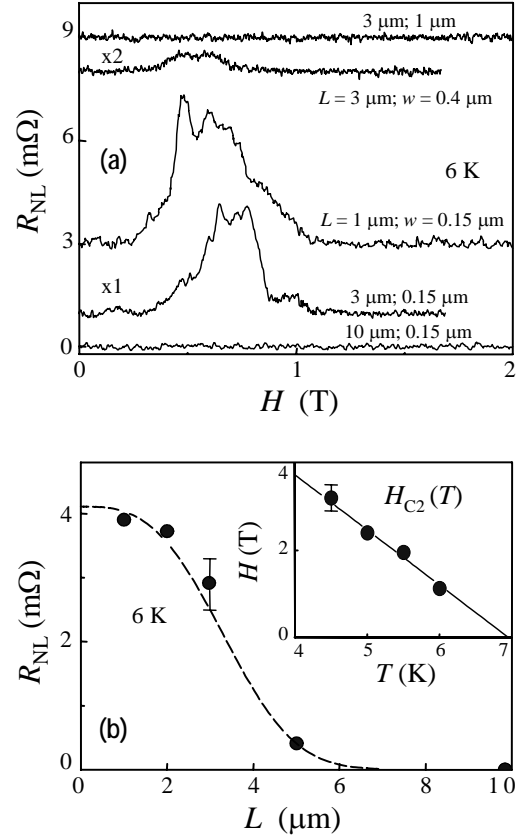

Figure 1                          Figure 2

**Figure 1.** Nonlocal resistance $R_{NL}$ as a function of applied field $H$ measured on a 150 nm wide wire at a distance of 1 μm between the current and voltage leads. Different curves are shifted vertically for clarity ($R_{NL}$ is always zero in the normal state). The inset shows an AFM image of the studied sample. The vertical wire in this image is referred to as central wire. Scale bar, 1 μm.

**Figure 2.** Dependence of $R_{NL}$ on length $L$ and width $w$ of the central wire. (a) Nonlocal resistance at $T = 6.0$ K for different wires (their $L$ and $w$ values are shown on the graph). Curves are shifted vertically for clarity. (b) Nonlocal resistance at its maximum value as a function of $L$ ($w = 150$ nm). The signal at 6.0K is also representative of the behaviour observed at lower $T$. The dashed line is a guide to the eye. The inset shows temperature dependence of the field corresponding to the disappearance of $R_{NL}$ (solid circles). The solid line is $H_{c2}(T)$ measured on macroscopic films.

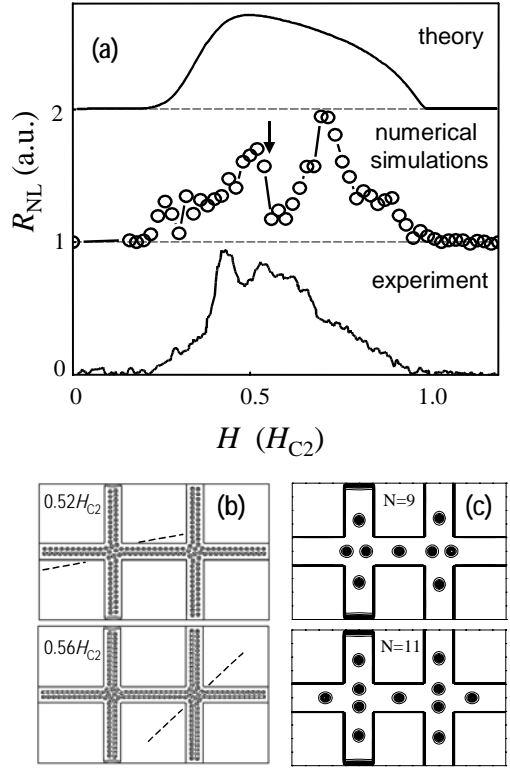

Figure 3

**Figure 3.** (a) Comparison of $R_{NL}(H)$ observed experimentally (lower curve; data of Fig. 1 at 6 K) with the nonlocal signal expected in the proposed 1D model (upper curve) and with results of our numerical analysis (middle curve). (b,c) Snapshots of vortex configurations corresponding to pronounced changes in mobility of 1D vortex matter. Dashed lines indicate the average orientation of vortex rows in the cross areas.